%% file: main.tex
\documentclass[conference]{IEEEtran}
\IEEEoverridecommandlockouts
\usepackage{cite}
\usepackage{amsmath,amssymb,amsfonts}
\usepackage{graphicx}
\usepackage{textcomp}
\usepackage{xcolor}
\usepackage{algorithm}
\usepackage{adjustbox}
\usepackage{algpseudocode}
\usepackage{booktabs}

\usepackage[]{footmisc}

\usepackage[numbers]{natbib}
\input{defs.tex}

\def\BibTeX{{\rm B\kern-.05em{\sc i\kern-.025em b}\kern-.08em
    T\kern-.1667em\lower.7ex\hbox{E}\kern-.125emX}}

\begin{document}

\title{Time-Series K-means in Causal Inference and Mechanism Clustering for Financial Data} 
\author{
\IEEEauthorblockN{1\textsuperscript{st}  Minheng Xiao}
\IEEEauthorblockA{\textit{dept. Integrated System Engineering} \\
\textit{Ohio State University}\\
Columbus, USA \\
minhengxiao@gmail.com}}

\maketitle

\begin{abstract}
\textbf{This paper examines the application of the time series $k$-means (TS-$k$-means) algorithm in the analysis of financial time series data. It employs dynamic time warping (DTW) as the metric and addresses the limitation of traditional clustering methods in capturing time dynamics. When combined with the additive noise Model Mixed Model (ANM-MM) architecture, this strategy not only enhances the accuracy of causal inference, but also enables clustering based on intrinsic generation mechanisms, thereby providing solid support for an in-depth understanding of complex financial time series data. The combination of TS-$k$-means and ANM-MM has been demonstrated to exhibit excellent performance in capturing fine causality and data clustering. Furthermore, it has been shown to possess significant advantages over traditional methods, while also demonstrating good stability and reliability on small-scale data sets. }
\end{abstract}

CCS CONCEPTS \\
Computing methodologies $\sim$ Machine learning $\sim$ Machine learning approaches $\sim$ Logical and relational learning $\sim$ Statistical relational learning


\section{Introduction}
Causal inference and causal relationship discovery between covariates (observations) is a foundational challenge across many domains, such as finance, social science and supply chain \citep{bo2024root}. Especially, understanding the underlying mechanisms and causal relationship of financial data is crucial for tasks such as risk management, portfolio optimization, and anomaly detection. Traditionally, clustering methods like $k$-means are utilized for grouping financial assets based on return patterns. However, traditional methods are often constrained by the limitations of static Euclidean distance, which is inadequate for fully capturing the dynamic changes, trends, and periodic time dependence inherent in financial time series. To fully analyze the temporal characteristics of financial data, flexible measures such as dynamic time distortion are necessary to transcend the concept of simple spatial distance. Some use deep learning-based methods \citep{dang2024realtime, article}, but the temporal dependences of financial data still can not be properly taken into account. Financial returns are not merely independent observations but are characterized by trends, seasonality, and varying volatility, making it imperative to adopt methods that respect the sequential nature of the data. Some machine learning methods, such as LightGMB \citep{Li_2024} and regression tree, are also static and can not deal with the time-series data appropriately.

Traditional approaches, such as LiNGAM \citep{shimizu2006linear}, PNL \citep{zhang2012identifiability} and IGCI \citep{janzing2010causal}, often assume that all observations are generated from a single, homogeneous causal model, which simplifies analysis but overlooks the complexity of real-world data. In practice, data is frequently collected from multiple environments or sources, each governed by distinct mechanisms that lead to heterogeneous causal relationships. This heterogeneity can render standard causal inference methods ineffective, as they fail to capture the underlying mixture of causal models. Recognizing this limitation, Hu et al. \cite{hu2018causal} proposed a novel approach by extending the Additive Noise Model \citep{hoyer2008nonlinear, reisach2024scale,zhao2024coresets} (ANM), $Y = f(X) + \varepsilon$, into a Mixture Model (ANM-MM) \citep{hu2018causal}. This model comprises multiple ANMs, each characterized by different functional forms and noise distributions, enabling the identification of diverse causal mechanisms within the data.

Their proposed framework integrates a Gaussian Process Partially Observable Model (GPPOM), which is a variant of the established GP-LVM \citep{lawrence2005probabilistic,maeda2024multi,tse2024causal}. This model estimates the latent parameters linked to each observation, enabling the identification of distinct causal mechanisms within the data. This framework addresses two core issues in financial time series analysis in an innovative manner. The first issue concerns the use of high-precision models to accurately identify the causal relationship between random variables, thereby deepening the insight into market dynamics. The second issue pertains to the use of cutting-edge clustering methods to accurately distinguish and cluster data observations of homologous generation mechanisms, which in turn allows for further analysis of market structure and potential opportunities.

In this work, we extend the ANM-MM framework to address the specific challenges posed by financial time series data. Traditional K-means clustering, commonly used for grouping financial assets, is limited by its reliance on Euclidean distance, which fails to capture the sequential dependencies in time series data. Financial returns, characterized by trends, volatility, and temporal patterns, require a more nuanced approach. To this end, we proposed a TS-$k$-means-based ANM-MM framework for financial transaction data, by replacing the static $k$-means algorithm with Time Series $k$-means, incorporating DTW as a distance metric. DTW enables the alignment of time series that may exhibit similar patterns with temporal shifts, The time series version of the $k$-means algorithm has been incorporated into the AM-MM framework, which has the effect of enhancing the resilience of financial time series data clustering. It is able to accurately capture market dynamics and sequence characteristics, reinforce the precision of causal reasoning and mechanism clustering, and establish a robust foundation for the comprehensive examination of financial markets.

$\mathbf{Paper\ Organization}$ Section~\ref{sec:2} provides a comprehensive definition of the symbols, an in-depth analysis of the background and original intention of the research, and a detailed examination of the key challenges and complexities inherent to financial time series analysis. Section~\ref{sec:3} presents a novel analytical framework that effectively integrates the strengths of TS-$k$-means clustering with the distinctive characteristics of financial data, offering a robust and sophisticated approach to the analysis of time series data and providing a powerful analytical tool for financial market analysis.

\section{Preliminaries \label{sec:2}}
The ANM hybrid model (ANM-MM) integrates the power of multiple additive noise models (ANM), whereby each model independently explores the potential causality of continuous variables \( X \) to \( Y \), flexibly responds to complex mechanisms, comprehensively captures multi-mode influences, and enhances the accuracy of causality analysis. This model is widely used in financial and biological information fields:

\[
Y = f(X; \theta) + \varepsilon,
\]

\begin{figure}
    \centering
    \includegraphics[width=1.3\linewidth, trim=1cm 2cm 0cm 1.5cm]{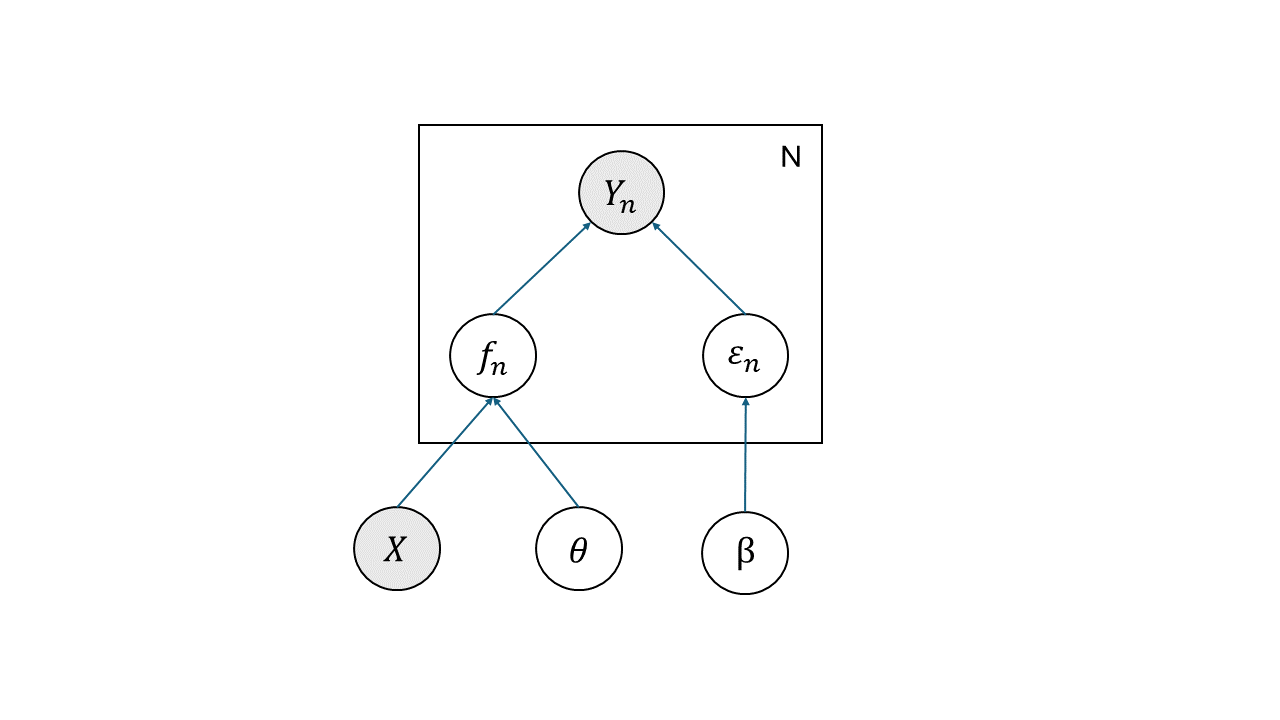}
    \caption{ANM-MM}
    \label{fig:ANM}
\end{figure}

In the ANN-mm model, \( X \) is the cause and \( Y \) is the effect, and the relationship between them is characterized by the nonlinear function \( f \) with the parameter \( \theta \), which is sampled from the finite set \( \Theta = \{\theta_1, \ldots, \theta_I\} \) discrete distribution to distinguish the causal models. Noise \( \varepsilon \) is independent of \( X \). All observations are generated by processes that share f but differ in \( \theta_i \), demonstrating the flexibility and breadth of the model in capturing complex causal relationships. This variation in parameters reflects the impact of external influences on the generating process in each independent trial. The process by which data is produced in the ANM-MM is depicted in Fig.\eqref{fig:ANM}.

This paper presents a new approach to causal inference within ANN-mm (artificial neural network hybrid modeling) by assessing the independence between a cause and model parameters to identify causality direction. Theoretical analysis shows that strong independence indicates the correct causal pathway, while dependence hints at a potential reverse relationship. Practically, measuring the degree of independence among variables aids in accurately identifying causal links. This method expands research in causal reasoning and improves the ANN-mm model's ability to analyze complex causal networks, highlighting its effectiveness and significance\cite{hu2018causal}. 

\section{Methodology \label{sec:3}}
The AAM-mm framework employs an innovative approach to parameter \( \theta \), which enables the direct and precise identification of diverse data generation mechanisms. Each distinct \( \theta \) is associated with a specific mechanism, thereby establishing a robust foundation for the identification of statistical models. This design not only optimizes the simplicity of the model architecture, but also significantly improves the accuracy and efficiency of data analysis. By relying on \( \theta \) sharing mechanism, AAM-mm enhances the functionality of data classification and aggregation, elucidates the interrelationship between data generation mechanisms, and introduces novel perspectives and practical tools for statistical modeling and data analysis \citep{hu2018causal}. Let \( \theta_n \) denote the parameter linked to each observation \( (x_n, y_n) \). \cite{hu2018causal} introduced a practical clustering structure within ANM-MM by allocating each \( \theta_n \) to one of the \( \theta_n \) components, with each component defined by a normal distribution \( \mathcal{N}(\mu_i, \sigma^2) \). They also showed that minimizing the likelihood-based loss function equates to applying the K-means algorithm to all \( \theta_n \). The ANN-mm architecture represents a novel integration of parameter estimation and K-means clustering technology. The approach begins with an accurate measurement of the parameters governing the generation of each observation value. These parameters are then subjected to direct cluster analysis, facilitating efficient data classification and fostering a deeper understanding of the data. Additionally, the ANN-mm architecture introduces novel perspectives and practical tools into the field of statistical modeling and data analysis. Given the temporal nature of financial time series data, it is hypothesized that for financial return time series, a time-series K-means algorithm based on DTW method \citep{kate2016using, hu2018clustering}, as proposed by \cite{kobylin2020time}, would be a more appropriate alternative to the standard K-means algorithm.

For two time series \( X_i \) of length \( n_i \), \( \{x_i\}_{i=1}^{n_i} \), and \( X_j \) of length \( n_j \), \( \{x_j\}_{j=1}^{n_j} \), the implementation of the DTW method can be outlined as follows \citep{kate2016using, hu2018clustering,rajput2024evaluating}. First, a distance matrix \( D = \{d_{i,j}\} \) is constructed, and then a transformation matrix \( D_{DTW} = \{r_{i,j}\} \) is designed. Each element is calculated with precision in order to transform the data and establish a foundation for comprehensive analysis

\[
r_{i,j} = d_{i,j} + \text{min} \left( D_{i-1,j}, D_{i-1,j-1}, D_{i,j-1} \right).
\]

In the processing of time series data, the determination of the optimal transformation path and the calculation of the \( D_{DTW} \) distance represent the fundamental steps in the measurement of series similarity. The traversal of the \( D_{DTW} \) matrix involves the careful selection of continuous elements with the objective of minimizing the total distance, thereby establishing the most suitable transformation path, \( W \). This process not only ensures the precise alignment between the sequences but also enables the accurate measurement of the similarity between the sequences through the use of the \( D_{DTW} \) distance. The incorporation of this powerful analytical tool into the field of time series analysis is a significant advancement \citep{kate2016using}:

\[
W = \{w_k\}_{k=1}^K, w_k = (i,j)_k, \text{max}(n_i,n_j) \leq K \leq (n_i + n_j),
\]

In the evaluation of time series similarity, \( K \) reflects the complexity of the underlying transformations. DTW distance formula synthesizes the distance and ek, quantifies the difference between sequences, and provides a quantitative index for analysis \citep{kate2016using, hu2018clustering}:

\[
DTW(X_i, X_j) = \text{min} \left( \frac{\sum_{k=1}^K d(w_k)}{K} \right).
\]

The mathematical validity of this approach, as per the setting proposed by \cite{hu2018causal}, will be investigated in forthcoming research.

\subsection{Algorithms}
Hu et al. \cite{hu2018causal} introduced GPPOM and HSIC \citep{gretton2005measuring,zhang2024statistical} to enhance the estimation accuracy of the model parameter \( \theta \). This framework combines causal inference and mechanism clustering algorithms to clarify the relationship between variables, remove noise, and reveal the data generation mechanism through mechanism clustering.

The \textbf{Gaussian Process Partially Observable Model (GPPOM)} extends GP-LVM \citep{lawrence2003gaussian} to handle scenarios where variables are only partially observable, within the context of the ANM Mixture Model. Leveraging kernel methods, GPPOM employs a nonlinear transformation from a latent space to the observed data realm via Gaussian Processes, effectively capturing intricate relationships between variables. Furthermore, GPPOM integrates HSIC aims to ensure the independence of latent variables from observable factors, which is essential for accurate causal inference and clustering of mechanisms. Through joint optimization, GPPOM estimates latent parameters that signify the underlying generative mechanisms, making it a robust tool for analyzing data originating from multiple causal processes.

HSIC, as detailed by \cite{gretton2005measuring}, is a widely adopted method based on reproducing kernel Hilbert space (RKHS) theory for assessing the dependence between random variables. Suppose we have a sample \( \mathcal{D} := \{(x_n, y_n)\}_{n=1}^N \) of size \( N \) drawn independently from the joint distribution \( P(X,Y) \). HSIC investigates whether \( X \) is independent of \( Y \) (denoted \( X \perp\!\!\!\perp Y \)). Formally, let \( \mathcal{F} \) and \( \mathcal{G} \) be RKHSs with universal kernels \( k \) and \( l \) on compact domains \( \mathcal{X} \) and \( \mathcal{Y} \), respectively. HSIC is defined as \( \text{HSIC}(P(X,Y), \mathcal{F}, \mathcal{G}) := \|\mathcal{C}_{xy}\|_{\text{HS}}^2 \), where \( \mathcal{C}_{xy} \) is the cross-covariance operator from RKHS \( \mathcal{G} \) to \( \mathcal{F} \), and \( \|\cdot\|_{\text{HS}} \) represents the Hilbert-Schmidt norm \citep{gretton2005measuring,nicola2024quantitative,fukumizu2004dimensionality}. \cite{gretton2005measuring} proved that under the conditions specified in \cite{gretton2005kernel}, \( \text{HSIC}(P(X,Y), \mathcal{F}, \mathcal{G}) = 0 \) if and only if \( X \perp\!\!\!\perp Y \). In practical scenarios, a biased empirical estimator of HSIC based on the sample \( \mathcal{D} \) is commonly used:

\[
\text{HSIC}_b(\mathcal{D}) = \frac{1}{N^2} \text{tr} ( \mathbf{KHLH} ),
\]

where \( [\mathbf{K}]_{ij} = k(x_i, x_j) \), \( [\mathbf{L}]_{ij} = l(y_i, y_j) \), \( \mathbf{H} = \mathbf{I} - \frac{1}{N} \mathbf{1} \mathbf{1}^\top \), and \( \mathbf{1} \) is a \( N \times 1 \) dimensional vector. In the G POM model, the optimization objective is expanded by introducing the HSIC term into its negative log-likelihood. This approach simultaneously optimizes the prediction accuracy, controls model complexity, and ensures the independence between variables, thereby achieving a more robust model performance:

\begin{align}
    \label{eq:obj}
\underset{\boldsymbol{\Theta}, \Omega}{\arg \min } \mathcal{J}(\boldsymbol{\Theta}) &= \underset{\boldsymbol{\Theta}, \Omega}{\arg \min }[-\mathcal{L}(\boldsymbol{\Theta} \mid \mathbf{X}, \mathbf{Y}, \Omega) \\
& + \lambda \log \mathrm{HSIC}_{\mathbf{b}}(\mathbf{X}, \boldsymbol{\Theta})],
\end{align}

where \( \lambda \) defines the significance of the HSIC component, and \( \Omega \) encompasses all hyperparameters, including \( \beta \) and the kernel parameters. For further technical details on GPPOM and the process of model and parameter estimation, see \citep{hu2018causal}. The following algorithm \ref{algo} provides a detailed approach for causal inference and mechanism clustering using time-series K-means. In Step 2, we implement time-series-based K-means utilizing DTW.

\begin{algorithm}[h!]
\label{algo}
\caption{Causal Inference and Mechanism Clustering with Time-Series K-means}
\begin{algorithmic}[1]
\State  Input dataset \( D = \{(x_n, y_n)\}_{n=1}^N \), specify independence parameter \( \lambda \) and number of clusters \( C \);
\State \textbf{Step 1: Causal Inference}
\State Standardize the observations for each random variable;
\State Initialize parameters \( \beta \) and kernel settings;
\State Optimize the objective in Eq. \eqref{eq:obj} for both directions, recording the HSIC values as \( \text{HSIC}_{X \rightarrow Y} \) and \( \text{HSIC}_{Y \rightarrow X} \);
\If{\( \text{HSIC}_{X \rightarrow Y} < \text{HSIC}_{Y \rightarrow X} \)}
    \State \textbf{Infer the causal direction as \( X \rightarrow Y \)};
\ElsIf{\( \text{HSIC}_{X \rightarrow Y} > \text{HSIC}_{Y \rightarrow X} \)}
    \State \textbf{Infer the causal direction as \( Y \rightarrow X \)};
\Else
    \State \textbf{No conclusive direction determined.}
\EndIf

\State \textbf{Step 2: Mechanism Clustering with Time-Series K-means}
\State Estimate \( \Theta \) by optimizing Eq. \eqref{eq:obj} in the determined causal direction;
\State Apply Time-Series K-means to \( \theta_n \) for \( n = 1, \dots, N \);
\State \textbf{Return} the cluster labels.
\end{algorithmic}
\end{algorithm}

\begin{figure}[h!]
    \centering
    \includegraphics[width=0.85\linewidth]{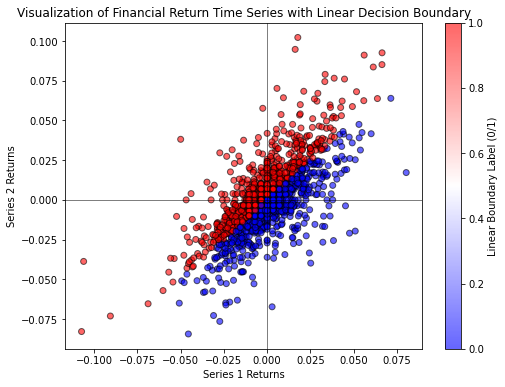}
    \caption{Linear clustering of data}
    \label{fig:linear}
\end{figure}
\section{Experiments \label{sec:4}}
\subsection{Simulation}
We first evaluated the performance of our TS-$k$-means approach through the simulation study. We refer to the setting in \cite{hu2018causal},
\begin{align*}
    X &\sim  \mathcal{U}(0,1)\\
    \theta_1 &\sim \mathcal{U}(1,1.1)\\
    \theta_2 &\sim \mathcal{U}(3,3.3)\\
    f_1 &= e^{-\theta_1 X}\\
    f_2 &= e^{-\theta_2 X}\\
    \varepsilon &\sim \mathcal{N}(0,0.05^2)\\
    Y &= f_i(X; \theta_i) + \varepsilon, \quad i \in \{1,2\}
\end{align*}
The performance of the model is assessed using avgARI \citep{hubert1985comparing,santri2024analysis,yadav2024refining}, which denotes the mean ARI computed across 100 experimental trials. ARI is a metric that quantifies the similarity between two clustering of data, with a correction for random groupings. The formula for ARI is as follows:

\[
\text{ARI} = \frac{\sum_{ij} \binom{n_{ij}}{2} - \left[ \sum_i \binom{a_i}{2} \sum_j \binom{b_j}{2} \right] / \binom{n}{2}}{\frac{1}{2} \left[ \sum_i \binom{a_i}{2} + \sum_j \binom{b_j}{2} \right] - \left[ \sum_i \binom{a_i}{2} \sum_j \binom{b_j}{2} \right] / \binom{n}{2}}
\]

Here, \( n_{ij} \) denotes the number of elements that are clustered in both \( U_i \) (from the clustering result) and \( V_j \) (from the ground truth). \( a_i \) high ARI value (ranging from -1 to 1) indicates a strong correspondence between the clustering result and the basic truth value. In this context, \( a_i \) and \( b_j \) represent the total number of elements in clusters \( U_i \) and \( V_j \), respectively, while \( b_j \) represents the total number of possible pairs.

\begin{table}[H]
\label{tab:1}
\centering
\caption{Results for simulated data}
\begin{tabular*}{\columnwidth}{@{\extracolsep{\fill}}ccccc|c}
\toprule
\textbf{Size} & \multicolumn{2}{c}{\textbf{avgARI}} & \multicolumn{2}{c|}{\textbf{HSIC}} & \\ \midrule
\textbf{n} & \textbf{$k$-means} & \textbf{TS-$k$-means} & $X \rightarrow Y$ & $Y \rightarrow X$ & \textbf{Direction} \\ \midrule
50  &  0.530&  \textbf{0.581}  &  2.101&  3.813& \textbf{X $\rightarrow$ Y} \\ 
100 & 0.791  &  \textbf{0.792}& 3.586 & 7.945 & \textbf{X $\rightarrow$ Y} \\ 
150 & 0.761 & 0.750 & 5.422 & 10.838 & \textbf{X $\rightarrow$ Y} \\ 
200 &  0.791& \textbf{0.792} & 6.592 & 17.157 & \textbf{X $\rightarrow$ Y} \\ 
252 & 0.687 & 0.678 & 6.506  &20.473  & \textbf{X $\rightarrow$ Y} \\ 
\bottomrule
\end{tabular*}
\label{tab:ex}
\end{table}
Table \eqref{tab:1} provides a comparison of ANM-MM K-means and TS-K-means on simulated datasets featuring diverse sample sizes (n = 50, 100, 150, 200, and 252). The evaluation metrics include the average Adjusted Rand Index (avgARI) and HSIC for assessing causal inference. As the size of the sample grows, the benefits of TS-K-means in comparison to K-means diminish, and K-means can sometimes perform comparably or better. In terms of causal inference, the HSIC values effectively identify the correct causal direction (X → Y) for all sample sizes, even as the data scale increases, maintaining consistent performance.

\begin{figure*}[h!]
    \centering
    \begin{tabular}{ccc}
        \includegraphics[scale=0.38]{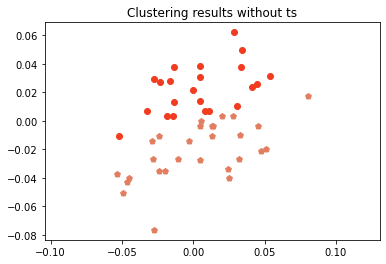} &
        \includegraphics[scale=0.38]{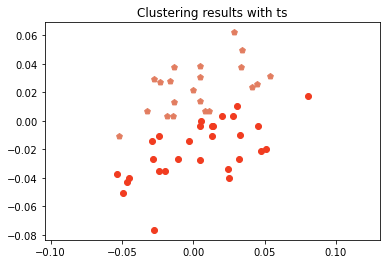} &
        \includegraphics[scale=0.38]{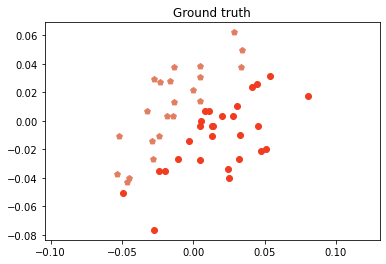} \\[-0.35cm]
        \includegraphics[scale=0.38]{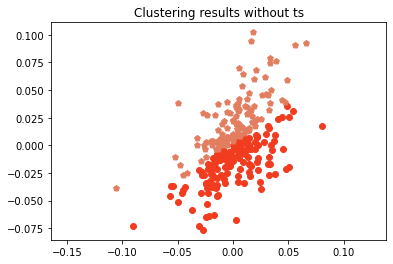}&
        \includegraphics[scale=0.38]{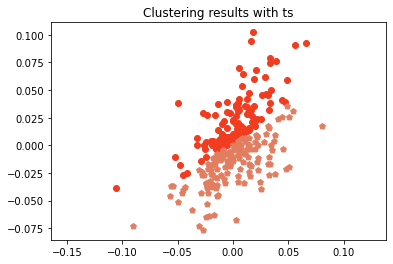}&
        \includegraphics[scale=0.38]{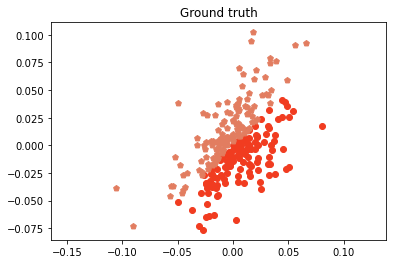}
    \end{tabular}
    \caption{ Ground truth and clustering results of different sample size (n = 50, 252)}
    \label{fig:clustering}
\end{figure*}

\subsection{Financial Data}
The dataset used in this work comprises daily stock returns of two prominent Hong Kong-listed companies: this study examines the market performance of Cheung Kong Holdings (0001.HK) and Sun Hung Kai Properties (0016.HK) from January 4, 2000, to June 17, 2005, using a causation database as the main data source for a comprehensive analysis \citep{mooij2016distinguishing}, with prices adjusted for dividends and stock splits. For days when the closing price was missing, The objective of linear interpolation is to estimate prices in a manner that ensures temporal alignment within a series and reduces the potential for bias in subsequent analysis. The daily returns for each stock were calculated using the formula \( X_t = \frac{P_t - P_{t-1}}{P_{t-1}} \), where \( P_t \) represents the closing price on day \( t \). This dataset provides a realistic context for analyzing causal relationships and as a constituent stock of the Hang Seng Real Estate Sub-Index, the financial returns of Sun Hung Kai Properties are subject to a time-dependent pattern that is significantly influenced by the performance of other major stocks in the index. The interaction between these two factors gives rise to a complex picture of market trends. The Fig. \eqref{fig:linear} shows the linearly clustered stock returns (we manually linearly clustered the data by $Y = X$ line). Obviously, there is positive correlation between two stock returns. By accurately classifying the returns into more meaningful clusters, we could improve risk assessment, and identify more precise causal relationships between the stocks. This refined classification would allow for better decision-making in financial strategies.

Tab. \eqref{tab:2} presents the performance of real financial data across varying sample sizes, with the smallest sample (n = 50) representing the first 50 trading days. When the sample size is small (n = 50), TS-$k$-means shows a noticeable improvement in avgARI compared to $k$-means, this also can be seen in Fig. \eqref{fig:clustering} , suggesting that even with limited data, accounting for temporal patterns leads to better clustering results. However, the HSIC values for causal inference indicate a strong preference for the incorrect direction (Y → X), which will change as the sample size enlarges. It becomes evident that an increase in sample size leads to more pronounced observations; the avgARI for both methods improves, but more importantly, the HSIC values gradually shift, indicating a correction in the inferred causal direction. By n = 150, the direction (X → Y) is consistently identified, underscoring how a larger window provides information on both clustering and causal inference. This shift highlights the causal relationship between two stocks may change as we expand the trading window.

\begin{table}[H]
\label{tab:2}
\centering
\caption{Results for real data}
\begin{tabular*}{\columnwidth}{@{\extracolsep{\fill}}ccccc|c}
\toprule
\textbf{Size} & \multicolumn{2}{c}{\textbf{avgARI}} & \multicolumn{2}{c|}{\textbf{HSIC}} & \\ \midrule
\textbf{n} & \textbf{$k$-means} & \textbf{TS-$k$-means} & $X \rightarrow Y$ & $Y \rightarrow X$ & \textbf{Direction} \\ \midrule
50  & 0.255 & \textbf{0.302}   & 1.087 & 5.801 & \textbf{X $\rightarrow$ Y} \\ 
100 & 0.330  & \textbf{0.353} & 0.668 & 0.746 & \textbf{X $\rightarrow$ Y} \\ 
150 & 0.391 & \textbf{0.406} & 0.907 & 0.627 & \textbf{Y $\rightarrow$ X} \\ 
200 & 0.487 & 0.487 & 1.255 & 0.514 & \textbf{Y $\rightarrow$ X} \\ 
252 & 0.543 & 0.543 & 1.298 & 0.527 & \textbf{Y $\rightarrow$ X} \\ 
\bottomrule
\end{tabular*}
\label{tab:ex}
\end{table}

\section{Conclusion}
In this research, we improved the traditional K-means clustering approach by incorporating TS-K-means into the ANM Mixture Model (ANM-MM) framework, with a focus on financial time series data. Our method utilizes DTW to effectively capture the temporal dependencies that are inherent in financial returns, which static clustering methods typically overlook. Through extensive simulations and analysis of real-world financial data, we demonstrated that TS-K-means consistently outperforms traditional K-means in terms of clustering accuracy, particularly for smaller datasets. Additionally, our approach maintains strong performance in causal inference, accurately identifying the correct causal direction across varying dataset sizes. These findings underscore the capability of TS-K-means to capture time-dependent patterns, thereby enhancing both clustering and causal inference in financial time series. We also observed that the causal relationship between variables may shift with changes in the trading window.

\section{Acknowledgement}
ChatGPT was employed to refine the wording and improve the readability of some sentences in this manuscript. Specifically, it was utilized for grammar and clarity checks in Sections III, as well as for enhancing the precision and readability of the conclusion section. The authors take full responsibility for the content and accuracy of this manuscript, including those sections where AI assistance was applied. 

\bibliographystyle{unsrtnat}
\bibliography{references}

\end{document}

%% file: defs.tex
\usepackage{amsmath}
\usepackage{amsfonts}
\usepackage{mathrsfs}
\usepackage{bbm}

\def\textbf#1{{\bfseries #1}}
\DeclareTextFontCommand{\textbf}{\bfseries}
\DeclareTextFontCommand{\textit}{\itshape}

\usepackage{amsthm}
\usepackage{etoolbox}
\theoremstyle{plain}